\documentclass[conference]{IEEEtran}
\IEEEoverridecommandlockouts
\usepackage{cite}
\usepackage{amsmath,amssymb,amsfonts}
\usepackage{algorithmic}
\usepackage{graphicx}
\usepackage{textcomp}
\usepackage{xcolor}
\def\BibTeX{{\rm B\kern-.05em{\sc i\kern-.025em b}\kern-.08em
    T\kern-.1667em\lower.7ex\hbox{E}\kern-.125emX}}
\usepackage{hyperref}
\usepackage{tikz}
\usepackage{tikzscale}
\usetikzlibrary{arrows,shapes,positioning}
\usetikzlibrary{calc,decorations.markings}

\DeclareSymbolFont{usualmathcal}{OMS}{cmsy}{m}{n}
\DeclareSymbolFontAlphabet{\mathcal}{usualmathcal}

\DeclareMathOperator{\expt}{\mathbb{E}} 

    \DeclareFontFamily{U}{wncy}{}
    \DeclareFontShape{U}{wncy}{m}{n}{<->wncyr10}{}
    \DeclareSymbolFont{mcy}{U}{wncy}{m}{n}
    \DeclareMathSymbol{\Sh}{\mathord}{mcy}{"58}

\begin{document}
  \newcommand{\ie}{i.\,e.}
  \newcommand{\eg}{e.\,g.}
  \tikzstyle{merge} = [draw, fill=yellow!20, minimum size=2em, align=center, circle]
\tikzstyle{layer} = [draw, fill=blue!20, minimum size=2em, align=center]
\tikzstyle{function} = [draw, fill=yellow!10, minimum size=2em, align=center]
\tikzstyle{neural} = [draw, fill=yellow!50, minimum size=2em, align=center]
\tikzstyle{filter} = [draw, fill=blue!10, minimum size=2em, align=center]
\tikzstyle{green} = [draw, fill=green!20, minimum size=2em, align=center]
\tikzstyle{input} = [draw=none, rounded corners=.2cm, fill=red!20, minimum size=2em, align=center]

\newcommand{\height}{}
\newcommand{\width}{}

  \title{Analysing Deep Learning-Spectral Envelope Prediction Methods for Singing Synthesis}

  \author{
    \IEEEauthorblockN{
      Frederik Bous
    }
    \IEEEauthorblockA{
      \textit{UMR STMS - IRCAM} \\
      \textit{CNRS, Sorbonne University}\\
      Paris, France \\
      frederik.bous@ircam.fr
    }
    \and
    \IEEEauthorblockN{
      Axel Roebel
    }
    \IEEEauthorblockA{
      \textit{UMR STMS - IRCAM} \\
      \textit{CNRS, Sorbonne University}\\
      Paris, France \\
      axel.roebel@ircam.fr
    }
  }

  \maketitle

  \begin{abstract}
    We conduct an investigation on various hyper-parameters
regarding neural networks
used to generate spectral envelopes for singing synthesis.
Two perceptive tests,
where the first compares two models directly
and the other ranks models with a mean opinion score,
are performed.
With these tests we show that
when learning to predict spectral envelopes,
2d-convolutions are superior
over previously proposed 1d-convolutions
and that predicting multiple frames
in an iterated fashion during training
is superior over injecting noise to the input data.
An experimental investigation
whether learning to predict a
probability distribution vs.\ single samples
was performed but turned out to be inconclusive.
A network architecture is proposed
that incorporates the improvements
which we found to be useful
and we show in our experiments
that this network produces better results
than other stat-of-the-art methods.

  \end{abstract}


  \section{introduction}
Singing synthesis is concerned with generating audio that sounds like a human singing voice from a musical description such as midi or sheet music. Compared with other musical instruments we observe that the human voice has one of the greatest varieties of possible sounds and the human ear is trained to distinguish the smallest differences in human voices. Human voice is among the first things a human learns and remains much apparent in our everyday life, and therefore almost everyone is a born \emph{expert} in perceiving human voice.

Compared with acoustic instruments,
singing not only incorporates melody and articulation,
but also text.
Compared with speech,
singing requires special treatment of the fundamental frequency $f_0$ as well as timing, which must be aligned to match melody and rhythm respectively.
However, due to its similarity to speech synthesis, more precisely text-to-speech (tts), many methods from tts may also be applied to singing synthesis. For years concatenative methods \cite{moulines1990pitch, hunt1996unit} dominated both fields \cite{ardaillon2017synthesis, bonada2016expressive, kenmochi2007vocaloid, gonzalvo2016recent}.
While these techniques yield fairly decent results,
they are inflexible and the underlying parametric speech models
usually treat all parameters independently,
which poses difficulties with coherency
of the parameters.
However, today fast computation on gpus
and large databases
allow us treating all parameters at once
in a single model
with neural networks
and they have already been successfully applied
to text-to-speech applications in the past years:

The system of \cite{zen2016fast} uses recurrent neural networks to model the statistic properties needed for their concatenative synthesis. WaveNet \cite{oord2016wavenet} goes further and models the raw audio, rather than concatenating existing audio or using a vocoder. Shortly after that, end-to-end systems like Tacotron \cite{shen2017natural} and Deep Voice \cite{ping2017deep} were developed which create raw audio from input on phoneme level or even character level. The authors of \cite{blaauw2017neural} used the architecture of \cite{oord2016wavenet} to learn input data for a parametric singing synthesizer.

While WaveNet processes data that is inherently one dimensional (i.\,e., raw audio), spectral envelopes are generated in \cite{blaauw2017neural}. There the input data is thus multidimensional, the authors use $60$ parameters to represent the spectral envelopes. This changes the nature of the data and former strong points of WaveNet may loose importance whereas some weaknesses may have a more significant impact. This has motivated our investigation into alternative network topologies and training strategies which finally has lead to an improved synthesis model.

We found that, contradicting the assumptions in \cite{blaauw2017neural}, 2d-convolutions yield better perceived audio while reducing the required number of trainable parameters. We also observe that learning by predicting multiple frames successively is superior to learning with additive noise at the input. A clear benefit from predicting parametric distributions rather than samples explicitly could not be found. As a result we propose our own network for predicting spectral envelopes.

The paper is structured as follows: we will first introduce our network in section \ref{sec:network} and discuss its differences to existing systems in section \ref{sec:differences}. The experimental setup will be explained in section \ref{sec:setup} and we present the results from our perceptive test in section \ref{sec:results}

\section{proposed network architecture}
\label{sec:network}
\begin{figure*}
  \centering
  \resizebox{\linewidth}{!}{%
    \includegraphics[]{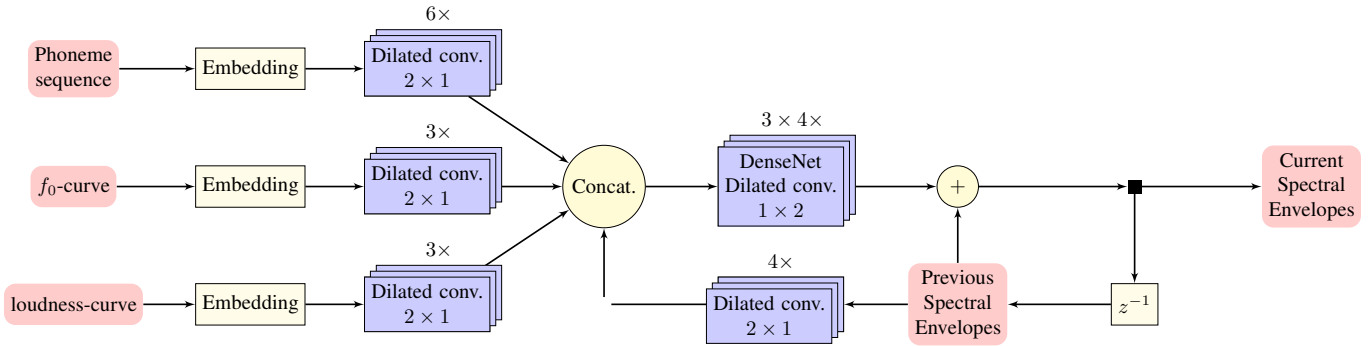}
  }
  \caption{Schematic layout of the network. Blue stacks denote stacks of layers, where $3 \times 4 \times$ means \emph{three stacks of four layers each}, $6 \times$ means \emph{one stack with six layers}. In each stack the dilation rate is doubled in each layer starting with a dilation rate of~$1$. The block $z^{-1}$ denotes a delay of one time step. Concatenation is done in the feature dimension.}
  \label{fig:architecture}
\end{figure*}
We aim to build a system for composers and professionals that wish to use synthetic singing voice in their compositions and applications. For us it is thus very important to keep a lot of flexibility in the system. While making application easy by automating obvious decisions, there should be as much ability to tweak all kinds of parameters as possible. Therefore end-to-end systems like Tacotron 2 \cite{shen2017natural}, where only the raw text is used as input and raw audio comes out as output and all other properties are only implicitly included in the model, if at all, do not fit our needs.

The role of the fundamental frequency $f_0$ is a very different in singing synthesis as compared to speech synthesis. In speech, the $f_0$-curve follows only few constraints but needs to be coherent with the other parameters. Learning it implicitly makes sense for end-to-end text-to-speech application as it does not carry much information, but coherence with other parameters is important. In singing, the $f_0$-curve is the parameter responsible for carrying the melody but it carries also musical style and emotion \cite{ardaillon2016expressive}. It is therefore important to model it explicitly, which can be achieved with, e.\,g., $B$-splines \cite{ardaillon2015multi}, to still be able to tweak it by hand to fit the needs of the particular application.

While systems like WaveNet \cite{oord2016wavenet} operate on raw audio, these architectures require very large datasets, which are currently not available for singing voice. This is for one due to less funding and for the other that recording proper singing requires even more work, as professional singers can not sing as long in one session as a professional speaker could speak.

In our application we use a vocoder model for singing voice synthesis. We use an improved version of the SVLN vocoder \cite{degottex2013mixed, huber2015glottal}, that is used to create singing voice from the modified parametric representation of the singing signal stored in a singing voice  database. In this context we aim to use a  neural network to provide  spectral envelopes that fit the local context (phoneme, F0, loudness) to the SVLN vocoder, that would then be responsible to generate the corresponding source signal.

\subsection{Input data}
Training data has been obtained from our own dataset of singing voice, which was originally created for a concatenative singing synthesizer. It consists of about 90 minutes of singing from a tenor voice. From this database we extract the spectral envelopes as well as the phoneme sequences, $f_0$-curves and loudness-curves as control parameters.

Phonemes were aligned with \cite{lanchantin2008automatic} and manually adjusted. Spectral envelopes are extracted from the audio with an improved version of the \emph{true envelope} method \cite{roebel2007cepstral}. The loudness curve is extracted by using a very simple implementation of the loudness model of Glasberg et al.\ \cite{glasberg2002model}, the $f_0$-curve is extracted by the pitch estimator of \cite{camacho2007swipe}. All data is given with a sampling rate of $200\mathrm{Hz}$ ($5\mathrm{ms}$ step size).

The spectral envelopes are represented by 60 log Mel-frequency spectral coefficients \cite{tokuda1994mel}, such that we can treat the spectral envelope sequence as a 2d spectrogram with Mel-spaced frequency bins. To obtain 60 bins but to keep a good resolution in the lower bins, we consider only frequencies up to $8\mathrm{kHz}$.

\subsection{Spectral Envelope Generation}
The spectral envelopes are generated by a recursive convolutional neural network. The network predicts one spectral envelope at a time by using the previous spectral envelopes as well as a window of the phoneme, $f_0$ and loudness values of previous, current and next time steps. We thus let the network to see a large context of control parameters from both future and past and thus allow it to create its own encoding.

The architecture is inspired by \cite{shen2017natural} and \cite{ping2017deep}. These systems use a neural network to create Mel-spectra, which are then converted to raw audio by either the Griffin-Lim algorithm or a vocoder. However, since we model the $f_0$ curve separately and do not encode it in the output, we can use a much simpler model.

Since all input parameters are given with the same rate, there is no need for attention. Only an encoding network of the control parameters, a pre-net for the previous spectral envelopes and a frame reconstruction network remain. In all parts we use blocks of dilated convolutions with exponentially growing dilation rate \cite{oord2016wavenet}, but additionally to dilated convolutions in time direction (as used in WaveNet and \cite{blaauw2017neural} and which we shall call $(2 \times 1)$-dilated convolutions) we also use dilated convolution in the frequency direction ($(1 \times 2)$-dilated convolutions).

We can summarise the architecture as follows (cf.\ Fig.\ \ref{fig:architecture}):
\begin{itemize}
  \item Input the envelopes from the last $2^{n_e}$ time steps and $2^{n_i}$ phoneme-values, $f_0$-values and loudness-values (where $n_i$ is different for each parameter) from a window of previous, current and next time steps around the current time step. The phonemes are mapped to $60$ frequency bins by an embedding layer, $f_0$-values and loudness-values are mapped to 60 frequencies by outer products.
  \item For each input parameter use a stack of $n_i$ dilated convolution layers of $(2\times1)$-convolutions and dilation rate $(2^{l}, 1)$ in layer $l$ (starting with $l=0$). No zero padding is done here. The convolution for the envelopes is deterministic, the other convolutions are symmetric.
  \item After the time-convolutions, the time-dimension is now one, while the frequency dimension remained $60$ for each input parameter. We concatenate all outputs from the time-convolutions along the feature dimension.
  \item The new frame is generated from the concatenation by several stacks of dilated convolution in the frequency-direction and with DenseNet skip-connections and bottleneck layers \cite{huang2017densely}. We use three stacks of four layers and use zero padding to keep the $60$ frequency dimensions.
  \item The final output is produced by a $(1\times1)$ convolution with one filter and adding the result to the previous frame. We thus only learn the difference from the previous frame to the next frame.
\end{itemize}

\subsection{Training}
The number of layers for the stacks of dilated convolutions in time-direction is $4$ for the spectral envelopes, $6$ for the phonemes and $3$ for both the $f_0$ and loudness.

We train the model using the adam optimizer \cite{kingma2014adam} with $\beta_1 = 0.9$ and $\beta_2 = 0.999$, just like in the original paper, but with an initial learning rate or $5 \cdot 10^{-4}$ and decay rate of $1 - 1 \cdot 10^{-5}$ per update (batch). We feed the network with minibatches consisting of 16 samples each chosen from random locations.

The loss is obtained as a simple mean squared error (mse) of the log amplitudes of the individual frequency bins. Other error functions like the mean absolute error or Sobolev norms (sums of $L^p$ norms and $L^p$ norms over its derivatives) were also considered but we found that results did not differ significantly.

\section{Differences with Existing Models}
\label{sec:differences}

\subsection{2d vs. 1d Convolutions}
\label{ssec:2d}
The authors of \cite{blaauw2017neural} claim that ``the translation invariance that 2d convolutions offer is an undesirable property for the frequency dimension''. Although in fact we do not expect to see every formant in each frequency bin with equal probability, formants can be found at different frequency locations. To be able to reduce the formants representation, we need to be able to shift the filters in time \emph{and} frequency.

To prove our claim, we build a 2d version of the WaveNet-style network from \cite{blaauw2017neural} and compare it to the original version to show that it yields in fact better audio.

The 2d version of \cite{blaauw2017neural} replaces $(2)$ dilated 1d convolutions with dilation rates of $2^l$ with $(2 \times 3)$ dilated 2d convolutions with dilation rate of $(2^l, 2^l)$. We reduce the number of filters dramatically so that we now have less trainable parameters (about one third) as compared to the original model but still more features per time step.

\subsection{Predicting Distributions}
It is common practice in prediction
to learn to predict distributions
rather than samples.
Distributions allow modelling data
that is uncertain or noisy.
In the case of WaveNet \cite{oord2016wavenet}
the system models a time series
that is a mix of a periodic signal
and coloured noise.
The coloured noise cannot be modelled
by a deterministic system
and therefore predicting a distribution
and sampling from it is necessary.

Reference \cite{blaauw2017neural} use the WaveNet architecture
to generate not audio, but spectral envelopes.
Their system predicts parameters of a constrained Gaussian mixture
to generate an independent parametric probability distribution
for each frequency bin of the spectral envelope.
However there are some very important differences
between raw audio and spectral envelopes:
raw audio (as modelled by WaveNet) has only one dimension per time step
while spectral envelopes are modelled (here) with $60$ frequency bins.
Raw audio is rapidly changing, contains oscillations and coloured noise,
while spectral envelopes are not oscillating, slowly changing and not noisy.

Since one time step of spectral envelopes
contains $60$ frequency bins,
it is impossible
to model all correlations of all frequency bins.
This is typically not necessary,
as correlations between frequency bins
that are far apart
can be assumed to be insignificant.
Nevertheless, there are correlations between neighbouring frequency bins
that cannot be neglected,
if the goal is to model the actual probability distribution
of the spectral envelopes.
Generating independent parametric distributions
for each frequency bin $F_i$
(as is done by \cite{blaauw2017neural})
must either assume that the frequency bins are independent
(which they are not)
or in fact yield an approximation of the true distribution
by the conditional expectations
$\tilde{F}_i = \expt(F_i|\{F_j:j\neq i\})$.
This is however the uninteresting part of the distribution.
The conditional expectation $\tilde{F}_i$ describes
the independent noise in each frequency band
while multiple possible positions of formants
are not modelled at all.

Since spectral envelopes are not noisy,
we believe that it is not necessary at all,
to predict probability distributions.
Our approach generates a spectral envelope directly.
This can be seen as generating the most probable sample
from the unknown (and unfeasible) distribution of the spectral envelopes.

%
%
%

\subsection{Stability by iterated prediction}
One problem with recursive models is stability. During prediction the error accumulates over time and once strayed too much from the path, there is no way to recover, because the system is in a state which it has never seen during training. It is also worth noting, that the envelopes do not change much during phonemes, but change more rapidly during a phoneme change.

To learn to make good predictions over a long time, a typical approach is to add noise to the input envelopes to simulate envelopes that have been previously predicted improperly or predicted properly, but were not contained in the training set. However the noise level needs to be very high and thus reduces the quality of the training data (Reference \cite{blaauw2017neural} suggests a noise level of $20\%$ of the value range). Instead, we enforce stability by iteratively predicting dozens of frames for each batch and applying the loss function to all predicted envelopes. This way we force the network to consider long term evolution and recover from prediction errors that are more likely to occur.

\section{Experimental Setup}
\newcommand{\BBone}{\textsf{BB1}}
\newcommand{\BBtwo}{\textsf{BB2}}
\newcommand{\MSE}{\textsf{MSE}}
\newcommand{\CGM}{\textsf{CGM}}
\newcommand{\iter}{\textsf{iter}}
\newcommand{\noise}{\textsf{noise}}
\label{sec:setup}
\begin{table}
  \caption{
    The different models that were trained
    for the perceptive tests.
  }
  \label{tab:models}
  \centering
  \begin{minipage}{\textwidth}
  \begin{tabular}{lllll}
    Name   & Architecture     & Conv. & Loss & Data Augmentation \\\hline
    \BBone & Blaauw \& Bonada & 1-d   & CGM\footnote{constrained Gaussian mixture from \cite{blaauw2017neural}}  & noise \\
    \BBtwo & Blaauw \& Bonada & 2-d   & CGM  & noise \\
    \MSE   & Bous \& Roebel   & 2-d   & MSE\footnote{mean squared error}  & iterated \\
    \CGM   & Bous \& Roebel   & 2-d   & CGM  & iterated \\
    \iter  & Bous \& Roebel   & 2-d   & MSE  & iterated \\
    \noise & Bous \& Roebel   & 2-d   & MSE  & noise \\
  \end{tabular}
  \end{minipage}
\end{table}
To support our claims from Section \ref{sec:differences} and to show that our network works well, we have conducted two perceptive tests with several different models: a direct comparison and a mos-test.

We train the networks on our singing database \cite{ardaillon2017synthesis} consisting of roughly $1000$ short phrases, and additional recordings of various pitches, loudnesses and crescendi, as well as short excerpts from real songs, from a single tenor voice, totalling about 90 minutes of singing voice. We split these recordings into training and testing files, where for each model we use the same training and testing files.

To regenerate the spectral envelopes with models that predict a probability distribution, we use the \emph{constrained Gaussian mixture} from \cite{blaauw2017neural} with a generation temperature of $\tau = 0$ to minimise sampling noise.

To obtain raw audio we resynthesize the testing files with the SVLN vocoder \cite{degottex2013mixed, huber2015glottal} by replacing the original spectral envelopes with the regenerated envelopes. We also include resynthesis with ground truth envelopes by resynthesizing the testing files without replacing the envelopes, thus resulting in a vocoder round trip. This procedure ensures that differences in the audio are exclusively due to differences in the  spectral envelopes that are used, and not to the use of the vocoder itself.

Two evaluate each of the proposed changes we perform a direct comparison of two models,
that differ only with respect to the single hyper-parameter subject to testing. Given our three
modifications we evaluate
\begin{itemize}
  \item the use of 2d versus 1d convolution by means of comparing our reimplementation of \cite{blaauw2017neural} and our modification as described in Section \ref{ssec:2d},
  \item the advantage of modelling predictions as probability distributions by means of comparing a model trained with mse-loss with another trained to maximise the log-likelihood of the distribution of predicted samples,
  \item iterated training by means of comparing a model that was trained with a single prediction with noise of $12\mathrm{db}$ standard deviation added to the input log-spectrum (the $12\mathrm{db}$ for the noise were found to work best among the values that were tested), and another model that trained recursively performing $24$ iterated predictions without any noise was added to the input.
\end{itemize}
To identify if the hyper-parameter is useful for overall quality, participants of our test were given the same phrase from both models and were asked to give a preference from $-3$ to $3$.

The mean opinion score has been measured by asking the participants to rank the given phrases on a scale from $1$ to $5$, where $1$ was the worst and $5$ was the best. Each participant was given the same phrase from all five models, but the phrases may differ for each participant. The models we used are summarised in Table \ref{tab:models}. For the mos test the following models were used: the two models from the 2d/1d comparison (\BBone\ and \BBtwo), the two models from the iterated vs.\ training with input noise comparison (\iter\ and \noise), and a resynthesis with ground truth envelopes.

The survey was carried out online. We received $31$ submissions from various backgrounds. Of those $31$ submissions, $9$ were from native French speakers.

\section{Results}
\label{sec:results}
Preferences of native French speakers are listed separately because the phrases were in French language. We can see that native French speakers were more critical (apparent in the mos test, cf.\ Table \ref{tab:mos}). This may be because native French speakers could additionally consider the pronunciation, and pronunciation may still not be as good (in the feed back it was actually mentioned that the singing voices seem to have kind of an accent).

\begin{table}
  \caption{%
    Perceptive test results of direct comparison.
    The preference is given towards the left model, \ie,
    a positive number implies a preference
    towards the first model.
    The $p$-value is the result of a one-sided $t$-test.
  }
  \label{tab:comp}
  \centering
\begin{tabular}{lrrrr}
  Comparison & Preference & $p$-Value & Preference & $p$-Value\\
   & (French) & (French) & (all) & (all)\\
  \hline
  \BBtwo\ vs.\ \BBone & $+1.44$ & $0.03\%$ & $+0.94$ & $0.00\%$\\
  \CGM\ vs.\ \MSE & $+0.00$ & $50.00\%$ & $+0.32$ & $3.73\%$\\
  \iter\ vs.\ \noise & $+0.78$ & $1.51\%$ & $+0.48$ & $0.26\%$\\
\end{tabular}

\end{table}

\begin{table}
  \caption{%
    Perceptive test results for mean opinion scores (mos)
    with $5\%$ confidence intervals.
  }
  \label{tab:mos}
  \centering
\begin{tabular}{lrr}
  Model & Mos (French) & Mos (all)\\
  \hline
  \iter & $3.15 \pm 0.38$ & $3.51 \pm 0.20$\\
  \noise & $3.11 \pm 0.33$ & $3.45 \pm 0.19$\\
  \BBone & $2.77 \pm 0.38$ & $2.96 \pm 0.24$\\
  \BBtwo & $3.11 \pm 0.46$ & $3.51 \pm 0.23$\\
  Ground truth & $3.56 \pm 0.53$ & $3.61 \pm 0.25$\\
\end{tabular}

\end{table}

\subsection{Comparison Test}
Table \ref{tab:comp} shows the results
from the comparison test.
The ``Preference'' column
contains the mean of the preference values
that were submitted towards the left model, \ie,
in the comparison \textsf{a} vs.\ \textsf{b}
positive values mean that \textsf{a} was preferred,
negative values mean that \textsf{b} was preferred.
The ``$p$-Value'' column
contains the $p$-value of the one sided Student-$t$-test, \ie,
the probability that,
the data was generated
under the alternative hypothesis
(``the right model was better or equal'').

There is a very clear preference
towards the use of 2-d convolutions
among both native French speakers
and all participants in total.
The $p$-value of $0.00\%$ actually means
that the $p$-value was below $0.005\%$,
which was rounded down to $0$.
Also a strong preference was given
towards the iterated training method.
No clear preference could be deduced
for the choice of loss function.
While a slight (and significant) preference
was given by all participants in total,
no preference was found among native French speakers.
Incidentally the preference values add up to zero,
however there were submissions with both
negative and positive preference.

\subsection{Mean Opinion Score Test}
Table \ref{tab:mos} shows the results
from the MOS test.
The given values are
the mean of the submitted scores
plus/minus half of the $5\%$ confidence interval
obtained by a two-sided Student-$t$-test.

The preferences are not as clear as compared to the comparison test.
While the relative preferences cannot be accepted
with a $p$-value of $5\%$ among native French speakers,
the preference against the state of the art \BBone\ 
is supported by the preferences of all participants.

The confidence intervals are rather large
due to the admittedly small number of participations.
The inferior conclusiveness of the MOS test
can be explained by its design:
During the MOS test
the participants were exposed to
the (almost) same recording five times.
While they might have heard some differences among the individual versions,
they were much more inclined
to put them in the same category
because they were still very similar,
than in the comparison test,
where they were explicitly asked
to favour one recording over the other.


\section{Conclusions}
In this paper we introduced a neural network architecture that is able to generate spectral envelopes for singing synthesis using a vocoder model. We showed in perceptive tests that the modifications we made with respect to the state-of-the-art method are useful in improving the perceptive result. In particular we showed that 2d convolutions are beneficial in modelling spectral envelopes and iteratively predicting multiple frames during training is superior to simply injecting noise at the input. An investigation whether predicting probability distributions rather than single samples was also carried out, but no benefit could be found when evaluating among native French speakers.

\section{Acknowledgments}
We like to thank Merlijn Blaauw for helpful discussions supporting our implementation of \cite{blaauw2017neural}

  \bibliographystyle{IEEEtran}
  \bibliography{IEEEabrv,../include/refs}

\end{document}